\documentclass[conference]{IEEEtran}
\IEEEoverridecommandlockouts
\usepackage{cite}
\usepackage{amsmath,amssymb,amsfonts}
\usepackage{algorithmic}
\usepackage{graphicx}
\usepackage{textcomp}
\usepackage{xcolor}
\usepackage{breqn}
\usepackage[xindy,acronym]{glossaries}
\usepackage{listings}
\usepackage{listings-owl}
\def\BibTeX{{\rm B\kern-.05em{\sc i\kern-.025em b}\kern-.08em
    T\kern-.1667em\lower.7ex\hbox{E}\kern-.125emX}}
\begin{document}


\newacronym{rag}{RAG}{Retrieval-Augmented Generation}
\newacronym{TV}{TV}{television}
\newacronym{DTTB}{DTTB}{digital television terrestrial broadcasting}
\newacronym{DVB}{DVB}{Digital Video Broadcast}
\newacronym{DVB-H}{DVB-H}{Digital Video Broadcast-Handheld}
\newacronym{ATSC}{ATSC}{Advanced Television System Committee}
\newacronym{ATSC-M/H}{ATSC-M/H}{Advanced Television System Committee - Mobile/Handheld}
\newacronym{IPTV}{IPTV}{Internet Protocol television}
\newacronym{IP}{IP}{Internet Protocol}
\newacronym{l1}{L1}{Layer 1}
\newacronym{l2}{L2}{Layer 2}
\newacronym{l3}{L3}{Layer 3}
\newacronym{cnn}{CNN}{Convolutional Neural Network}
\newacronym{ann}{ANN}{Artificial Neural Network}
\newacronym{scef}{SCEF}{Service Capability Exposure Function}
\newacronym{zipnet}{ZipNet}{Zipper Network}
\newacronym{rrm}{RRM}{Radio Resource Management}
\newacronym{fps}{FPS}{Frames Per Second}
\newacronym{qa}{QA}{Question Answering}
\newacronym{ar}{AR}{Augmented Reality}

\newacronym{cdl}{CDL}{Clustered Delay Line}
\newacronym{tdl}{TDL}{Tapped Delay Line}

\newacronym{HR}{HR}{Human Resources}

\newacronym{tot}{ToT}{Tree-of-Thoughts}
\newacronym{got}{GoT}{Graph-of-Thoughts}

\newacronym{DGM}{DGM}{Deep Generative Model}
\newacronym{PGM}{PGM}{Probabilistic Graphical Model}
\newacronym{VAE}{VAE}{Variational autoencoder}
\newacronym{ARM}{ARM}{Autoregressive model}
\newacronym{NF}{NF}{Normalizing flows}
\newacronym{PDDL}{PDDL}{Planning Domain Definition Language}
\newacronym{RL}{RL}{Reinforcement Learning}
\newacronym{react}{ReAct}{Reasoning and Acting}
\newacronym{pal}{PAL}{Program-aided Language Models}

\newacronym{CSI}{CSI}{Channel State Information}
\newacronym{ml}{ML}{Machine Learning}
\newacronym{bss}{BSS}{Business Support System}
\newacronym{nlp}{NLP}{Natural Language Processing}
\newacronym{1g}{1G}{first generation of mobile networks}
\newacronym{2g}{2G}{second generation of mobile networks}
\newacronym{2.5g}{2.5G}{Transitional 2.5 generation of mobile networks}
\newacronym{3g}{3G}{third generation of mobile networks}
\newacronym{4g}{4G}{fourth generation of mobile networks}
\newacronym{5G}{5G}{fifth generation of mobile networks}
\newacronym{5g}{5G}{fifth generation of mobile networks}
\newacronym{6g}{6G}{sixth generation of mobile networks}
\newacronym{nlg}{NLG}{natural language generation}
\newacronym
[
  longplural={Large Language Models}
]
{llm}{LLM}{Large Language Model}
\newacronym
[
  longplural={Bidirectional Encoder Representations from Transformers}
]
{bert}{BERT}{Bidirectional Encoder Representations from Transformer}
\newacronym
[
  longplural={Universal Sentence Encoder}
]
{use}{USE}{Universal Sentence Encoder}
\newacronym
[
  longplural={Radio Base Stations}
]
{rbs}{RBS}{Radio Base Station}
\newacronym
[
  longplural={Natural Language Understanding}
]
{nlu}{NLU}{Natural Language Understanding}
\newacronym
[
  longplural={Network Functions}
]
{nf}{NF}{Network Function}
\newacronym
[
  longplural={Generative Pretrained Transformers}
]
{gpt}{GPT}{Generative Pretrained Transformer}

\newacronym
[
  longplural={Knowledge Bases}
]
{kb}{KB}{Knowledge Base}

\newacronym{palm}{PaLM}{Pathways Language Model}
\newacronym{llama}{LLAMA}{Large Language Model Meta AI}

\newacronym{cot}{CoT}{Chain-of-Thought}

\newacronym
[
  longplural={Generative Pretrained Transformers}
]
{rouge}{ROUGE}{Recall-Oriented Understudy for Gisting Evaluation}
\newacronym
[
  longplural={Deep Belief Networks}
]
{dbn}{DBN}{Deep Belief Network}
\newacronym
[
  longplural={Boltzmann Machines}
]
{bm}{BM}{Boltzmann Machine}
\newacronym
[
  longplural={Variational Autoencoders}
]
{vae}{VAE}{Variational Autoencoder}
\newacronym
[
  longplural={Long-Short Term Memory networks}
]
{lstm}{LSTM}{Long-Short Term Memory network}
\newacronym
[
  longplural={Probability Density Functions}
]
{pdf}{PDF}{Probability Density Function}
\newacronym
[
  longplural={Recurrent Neural Networks}
]
{rnn}{RNN}{Recurrent Neural Network}
\newacronym
[
  longplural={Generative Adversarial Networks}
]
{gan}{GAN}{Generative Adversarial Network}
\newacronym
[
  longplural={Restricted Boltzmann Machines}
]
{rbm}{RBM}{Restricted Boltzmann Machine}
\newacronym
[
  longplural={Digital Twins}
]
{dt}{DT}{Digital Twin}

\newacronym{itu}{ITU}{International Telecommunication Union}
\newacronym{etsi}{ETSI}{European Telecommunication Standards Institute}
\newacronym{api}{API}{Application Program Interface}
\newacronym{ue}{UE}{User Equipment}
\newacronym{PC}{PC}{personal computer}
\newacronym{RAN}{RAN}{radio access network}
\newacronym{CN}{CN}{core network}
\newacronym{MS}{MS}{mobile station}
\newacronym{soa}{SoA}{state of the art}

\newacronym{genai}{GenAI}{Generative AI}

\newacronym{ITU-R}{ITU-R}{International Telecommunications Union - Radiocommunication Sector}
\newacronym{IMT-Advanced}{IMT-Advanced}{International Mobile Telecommunications Advanced}
\newacronym{4G}{4G}{fourth-generation of mobile phone communications and Internet access technology}

\newacronym{rlhf}{RLHF}{Reinforcement Learning with Human Feedback}
\newacronym{3gpp}{3GPP}{Third Generation Partnership Project}
\newacronym{GSM}{GSM}{Global System for Mobile Communications}
\newacronym{UMTS}{UMTS}{Universal Mobile Telecommunications System}
\newacronym{HSPA}{HSPA}{High Speed Packet Access}
\newacronym{lte}{LTE}{Long-Term Evolution}
\newacronym{lte-a}{LTE-A}{Long-Term Evolution Advanced}

\newacronym{m2m}{M2M}{Machine-to-machine}

\newacronym{OSS}{OSS}{Operations Support System}

\newacronym{tmn}{TMN}{Telecommunications Management Network}

\newacronym{e-UTRAN}{e-UTRAN}{evolved Universal Terrestrial Radio Access Network}
\newacronym{eNB}{eNB}{e-UTRAN NodeB}
\newacronym{gNB}{gNB}{gNodeB}
\newacronym{EPC}{EPC}{Evolved Packet Core}

\newacronym{MBMS}{MBMS}{Multimedia and Broadcast Multicast Service}
\newacronym{eMBMS}{eMBMS}{Evolved MBMS}
\newacronym{SFN}{SFN}{single-frequency network}
\newacronym{MBSFN}{MBSFN}{MBMS single-frequency network}
\newacronym{BM-SC}{BM-SC}{Broadcast/Multicast Service Center}
\newacronym{MBMS GW}{MBMS GW}{MBMS Gateway}
\newacronym{MME}{MME}{Mobility Management Entity}
\newacronym{MCE}{MCE}{Multi-cell/multicast Coordinating Entity}
\newacronym{SYNC}{SYNC}{synchronization}
\newacronym{MCCH}{MCCH}{Multicast Control Channel}
\newacronym{MTCH}{MTCH}{Multicast Traffic Channel}
\newacronym{MCH}{MCH}{Multicast Channel}
\newacronym{PMCH}{PMCH}{Physical Multicast Channel}
\newacronym{PDSCH}{PDSCH}{Physical Downlink Shared Channel}
\newacronym{tmforum}{TMForum}{TeleManagement Forum}

\newacronym{sop}{SOP}{Sentence Order Prediction}
\newacronym{fid}{FID}{Fréchet Inception Distance}
\newacronym{is}{IS}{Inception Score}
\newacronym{mlm}{MLM}{Masked Language Modeling}
\newacronym{nsp}{NSP}{Next Sentence Prediction}
\newacronym{blue}{BLUE}{Bilingual Evaluation Understudy}
\newacronym{cer}{CER}{Concept Error Rate}

\newacronym{hss}{HSS}{Home Subscriber Server}

\newacronym{sft}{SFT}{Supervised Fine-Tuning}
\newacronym{gpu}{GPU}{Graphics Processing Unit}
\newacronym{IEEE}{IEEE}{Institute of Electrical and Electronics Engineers}
\newacronym{WiMAX}{WiMAX}{Worldwide Interoperability for Microwave Access}
\newacronym{ASN}{ASN}{access service network}
\newacronym{ASN-GW}{ASN-GW}{ASN gateway}
\newacronym{CSN}{CSN}{Connectivity Service Network}
\newacronym{oran}{O-RAN}{Open Radio Access Network}
\newacronym{ric}{RIC}{Radio Intelligent Controller}
\newacronym{rt}{RT}{Real-Time}
\newacronym{uav}{UAV}{Unidentified Aerial Vehicle}
\newacronym{v2x}{V2X}{Vehicle-To-Everything}

\newacronym{nwdaf}{NWDAF}{Network Data Analytics Function}
\newacronym{pcf}{PCF}{Policy Control Function}
\newacronym{mtlf}{MTLF}{Model Training Logical Function}
\newacronym{anlf}{ANLF}{Analytics Logical Function}

\newacronym{PA}{PA}{power amplifier}
\newacronym{ecgan}{ECGAN}{Enhanced Capsule Generation Adversarial Network}

\newacronym{NI}{NI}{National Instruments}

\newacronym{TDD}{TDD}{time-division duplex}
\newacronym{FDD}{FDD}{frequency-division duplex}
\newacronym{UDP}{UDP}{User Datagram Protocol}
\newacronym{APP}{APP}{application}
\newacronym{mac}{MAC}{medium access control}
\newacronym{phy}{PHY}{physical}
\newacronym{RLC}{RLC}{radio link control}
\newacronym{sdap}{SDAP}{service data adaptation protocol}
\newacronym{FIFO}{FIFO}{first-in first-out}
\newacronym{CRC}{CRC}{cyclic redundancy check}
\newacronym{SAP}{SAP}{service access point}
\newacronym{FEC}{FEC}{forward error correction}
\newacronym{IF}{IF}{intermediate frequency}
\newacronym{RF}{RF}{radio frequency}
\newacronym{mimo}{MIMO}{multiple-input and multiple-output}
\newacronym{MCS}{MCS}{modulation and coding scheme}

\newacronym{SPC}{SPC}{superposition coding}
\newacronym{SVC}{SVC}{Scalable Video Coding}
\newacronym{GM}{GM}{generic multicasting}
\newacronym{SCM}{SCM}{superposition coded multicasting}
\newacronym{SIC}{SIC}{successive interference cancellation}

\newacronym{st}{ST}{secondary transmitter}
\newacronym{pt}{PT}{primary transmitter}
\newacronym{sr}{SR}{secondary receiver}
\newacronym{pr}{PR}{primary receiver}
\newacronym{su}{SU}{secondary user}
\newacronym{pu}{PU}{primary user}

\newacronym{awgn}{AWGN}{additive white Gaussian noise}

\newacronym{cdf}{CDF}{cumulative density function}
\newacronym{ccdf}{CCDF}{complementary CDF}
\newacronym{iid}{i.i.d.}{independent and identically distributed}
\newacronym{rf}{RF}{radio frequency}
\newacronym{hbf}{HBF}{Hybrid Beamforming}

\newacronym{dd}{DD}{Device-to-Device}
\newacronym{ddu}{DDU}{Device-to-Device user}
\newacronym{dds}{DDS}{Device-to-Device system}
\newacronym{ddt}{DT}{DDU transmitter}
\newacronym{ddr}{DR}{DDU receiver}

\newacronym{bs}{BS}{base station}
\newacronym{bsu}{BSU}{base station associated user}
\newacronym{bsas}{BSAS}{base station associated system}
\newacronym{bst}{BT}{BSU transmitter}
\newacronym{bsr}{BR}{BSU receiver}

\newacronym{epg}{EPG}{energy per goodbit}
\newacronym{mepg}{MEPG}{modified energy per goodbit}
\newacronym{ee}{EE}{energy efficiency}
\newacronym{se}{SE}{spectral efficiency}

\newacronym{wrt}{w.r.t.}{with respect to}

\newacronym{kkt}{KKT}{Karush-Kuhn-Tucker}
\newacronym{al}{AL}{Active Learning}
\newacronym{admm}{ADM}{Alternating Directing Method}
\newacronym{cr}{CR}{cognitive radio}
\newacronym{ssi}{SSI}{soft-sensing information}
\newacronym{csi}{CSI}{Channel State Information}
\newacronym{qsi}{QSI}{queue state information}
\newacronym{el}{EL}{enhancement layer(s)}
\newacronym{snr}{SNR}{signal-to-noise ratio}

\newacronym{NAL}{NAL}{network abstraction layer}
\newacronym{QP}{QP}{quantization parameter}

\newacronym{ofdm}{OFDM}{orthogonal frequency-division multiplexing}
\newacronym{ofdma}{OFDMA}{orthogonal frequency-division multiple access}
\newacronym{tdma}{TDMA}{time division multiple access}

\newacronym{PUSC}{PUSC}{partial usage of the subchannels}
\newacronym{CFO}{CFO}{carrier frequency offset}
\newacronym{I/Q}{I/Q}{in-phase and quadrature-phase}
\newacronym{ASK}{ASK}{amplitude-shift keying}
\newacronym{PSK}{PSK}{phase-shift keying}
\newacronym{BPSK}{BPSK}{binary phase-shift keying}
\newacronym{QPSK}{QPSK}{quadrature phase-shift keying}
\newacronym{QAM}{QAM}{quadrature amplitude modulation}
\newacronym{PSNR}{PSNR}{peak signal-to-noise ratio}
\newacronym{PELR}{PELR}{packet error and loss rate}

\newacronym{kNN}{\textit{k}-NN}{\textit{k}-nearest neighbor algorithm}
\newacronym{SVM}{SVM}{support vector machines}
\newacronym{nn}{NN}{neural network}
\newacronym{NN}{NN}{neural network}
\newacronym{dnn}{DNN}{deep neural network}
\newacronym{RBF}{RBF}{radial basis function}
\newacronym{RMSE}{RMSE}{root mean squared error}
\newacronym{mse}{MSE}{mean squared error}
\newacronym{lmse}{LMSE}{linear mean square-error estimator}

\newacronym{R2}{$R^2$}{coefficient of determination}

\newacronym{KAUST}{KAUST}{King Abdullah University of Science and Technology}
\newacronym{GSA}{GSA}{Global mobile Suppliers Association}

\newacronym{VoD}{VoD}{video on demand}
\newacronym{HEVC}{HEVC}{High Efficiency of Video Coding}
\newacronym{DASH}{DASH}{Dynamic Adaptive Streaming over HTTP}

\newacronym{PUT}{PUT}{people using television}

\newacronym{ADTVS}{ADTVS}{Audience Driven live TV Scheduling}

\newacronym{arq}{ARQ}{automatic repeat request}

\newacronym{harq}{HARQ}{hybrid automatic repeat request}

\newacronym{sdp}{SDP}{semi-definite programming}

\newacronym{tcp}{TCP}{transmission control protocol}

\newacronym{e2e}{E2E}{end-to-end}

\newacronym{ran}{RAN}{radio access network}
\newacronym{cran}{CRAN}{cloud radio access network}
\newacronym{udcran}{UD-CRAN}{ultra-dense CRAN}
\newacronym{dran}{DRAN}{distributed radio access network}
\newacronym{hcran}{H-CRAN}{hybrid cloud radio access network}
\newacronym{hetnet}{HetNet}{heterogeneous network}
\newacronym{vcran}{V-CRAN}{virtualized CRAN}
\newacronym{ecran}{E-CRAN}{edge-CRAN}
\newacronym{hvcran}{H-VCRAN}{hybrid-virtualized CRAN}

\newacronym{bbu}{BBU}{baseband processing unit}
\newacronym{rrh}{RRH}{remote radio head}
\newacronym{ru}{RU}{radio unit}
\newacronym{rs}{RS}{remote site}
\newacronym{cs}{CS}{central site}

\newacronym{rru}{RRU}{radio resource unit}
\newacronym{rb}{RB}{resource block}
\newacronym{hpn}{HPN}{high-power node}
\newacronym{lpn}{LPN}{low-power node}
\newacronym{mabs}{MaBS}{macro basestation}

\newacronym{comp}{CoMP}{coordinated multi-point}
\newacronym{ranaas}{RANaaS}{RAN-as-a-Service}

\newacronym{rof}{RoF}{radio over fiber}
\newacronym{wdm}{WDM}{Wavelength Division Multiplexing}
\newacronym{dls}{DLS}{distributed large scale}

\newacronym{qos}{QoS}{quality of service}
\newacronym{qoe}{QoE}{quality of experience}
\newacronym{qee}{QEE}{quality of energy-efficiency}

\newacronym{gg}{GG}{group-to-group}
\newacronym{ht}{HT}{hyper-transceiver}

\newacronym{fh}{FH}{fronthaul}
\newacronym{dl}{DL}{downlink}
\newacronym{ul}{UL}{uplink}

\newacronym{cp}{CP}{Cell-Processing}
\newacronym{up}{UP}{User-Processing}

\newacronym{sla}{SLA}{Service-Level Agreement}

\newacronym{kr}{KR}{knowledge representation}

\newacronym{co}{CO}{center office}

\newacronym{krr}{KRR}{Knowledge Representation and Reasoning}
\newacronym{du}{DU}{digital unit}
\newacronym{lc}{LC}{Line-Card}

\newacronym{onu}{ONU}{optical network unit}
\newacronym{olt}{OLT}{optical line terminal}
\newacronym{osw}{OSW}{optical switch}

\newacronym{es}{ES}{ethernet switch}

\newacronym{ppp}{PPP}{Poisson point process}

\newacronym{mppp}{MPPP}{marked Poisson point process}

\newacronym{sinr}{SINR}{signal to noise and interference ratio}

\newacronym{sir}{SIR}{signal to interference ratio}

\newacronym{mbs}{MBS}{macro basestation}
\newacronym{ap}{AP}{access point}
\newacronym{fap}{FAP}{femto-cell access point}
\newacronym{sap}{SAP}{small-cell access point}
\newacronym{iot}{IoT}{Internet of Things}
\newacronym{ti}{TI}{Tactile Internet}
\newacronym{ntn}{NTN}{Non-Terrestrial Network}
\newacronym{lsm}{LSM}{linear scalarizing method}

\newacronym{lp}{LP}{Low-Priority}
\newacronym{hp}{HP}{High-Priority}
\newacronym{lpu}{LPU}{Low-Priority user}
\newacronym{hpu}{HPU}{High-Priority user}
\newacronym{lps}{LPS}{Low-Priority system}
\newacronym{hps}{HPS}{High-Priority system}

\newacronym{ietf}{IETF}{Internet Engineering Task Force}

\newacronym{ttm}{TTM}{time to market}
\newacronym{udn}{UDN}{ultra-dense network}

\newacronym{capex}{CAPEX}{capital expenditure}
\newacronym{opex}{OPEX}{operational expenditure}

\newacronym{cpri}{CPRI}{common public radio interface}
\newacronym{otn}{OTN}{optical transport network}
\newacronym{pon}{PON}{passive optical network}
\newacronym{twdm}{TWDM}{time and wavelength division multiplexing}

\newacronym{ec}{EC}{Edge-Cloud}
\newacronym{cc}{CC}{Central-Cloud}

\newacronym{mmw}{m-Wave}{Milli-Meter wave}

\newacronym{gops}{GOPS}{giga operation per second}
\newacronym{mops}{MOPS}{mega operation per second}

\newacronym{ipr}{IP}{Intellectual Property}

\newacronym{ip}{IP}{internet protocol}
\newacronym{rlc}{RLC}{radio link control}
\newacronym{pdcp}{PDCP}{packet data convergence protocol}

\newacronym{mno}{MNO}{mobile network operator}
\newacronym{prb}{PRB}{physical resource block}
\newacronym{mi}{MI}{modulation index}

\newacronym{wifi}{WiFi}{wireless local area network}
\newacronym{cpu}{CPU}{central processing unit}
\newacronym{vcpu}{VCPU}{virtual CPU}
\newacronym{vm}{VM}{virtual machine}

\newacronym{urs}{UrS}{user requested service}

\newacronym{npcomplete}{NP-complete}{nondeterministic polynomial-time complete}

\newacronym{rsf}{RSF}{radio sub-frame}
\newacronym{siso}{SISO}{single-input single-output}
\newacronym{ram}{RAM}{random access memory}
\newacronym{xr}{XR}{Extended Reality}

\newacronym{nef}{NEF}{Network Exposure Function}

\newacronym{vr}{VR}{Virtual Reality}
\newacronym{agv}{AGV}{Automated Guided Vehicle}

\newacronym{rssi}{RSSI}{Received Signal Strength Indicator}

\newacronym{mec}{MEC}{mobile edge computing}
\newacronym{co2}{CO$_{2}$}{carbo dioxide}

\newacronym{cfp}{CFP}{communication function processing}
\newacronym{ptp}{PTP}{precision time protocol}

\newacronym{mdt}{MDT}{Model Drive Test}

\newacronym{voip}{VoIP}{voice over Internet protocol}
\newacronym{sdn}{SDN}{Software Defined Network}

\newacronym{da}{DA}{data analytics}

\newacronym{kpi}{KPI}{key performance indicator}

\newacronym{noc}{NOC}{Network Operations Centre}
\newacronym{fso}{FSO}{Field Service Operations}

\newacronym{fsmc}{FSMC}{finite state markov chain}

\newacronym{nr}{NR}{new radio}
\newacronym{gnbcu}{gNB-CU}{gNB central unit}
\newacronym{gnbdu}{gNB-DU}{gNB distributed unit}
\newacronym{ecpri}{eCPRI}{common public radio interface}
\newacronym{fl}{FL}{federated learning}
\newacronym{rsrq}{RSRQ}{Reference Signal Received Quality}
\newacronym{rsrp}{RSRP}{Reference Signal Received Power}
\newacronym{urllc}{URLLC}{ultra-reliable low-latency communications}
\newacronym{embb}{eMBB}{enhanced mobile broadband}

\newacronym{mae}{MAE}{modified autoencoder}
\newacronym{mtc}{MTC}{machine type communication}
\newacronym{mmtc}{mMTC}{massive machine type communication}
\newacronym{pca}{PCA}{principal component analysis}

\newacronym{cps}{CPS}{cyber-physical system}
\newacronym{gnb}{gNB}{gNodeB}
\newacronym{ref}{REF}{reliability enhancement feature}
\newacronym{nfo}{NFO}{network level feature orchestrator}
\newacronym{dc}{DC}{data center}
\newacronym{vnf}{VNF}{virtual network function}
\newacronym{nssmf}{NSSMF}{Network Slice Subnet Management Function}
\newacronym{ai}{AI}{Artificial Intelligence}
\newacronym{rl}{RL}{Reinforcement Learning}
\newacronym{ddpg}{DDPG}{deep deterministic policy gradient}
\newacronym{dqn}{DQN}{deep Q-networks}
\newacronym{sac}{SAC}{soft actor-critic}
\newacronym{a2c}{A2C}{advantage actor-critic}
\newacronym{td3}{TD3}{twin delayed deep deterministic policy gradient algorithm}
\newacronym{poc}{PoC}{proof of concept}
\newacronym{cvae}{CVAE}{Conditional Variational AutoEncoder}
\newacronym{oam}{OAM}{Operation, Administration, and Maintenance}

\newcommand*\myglsentry[1]{%
  \protect\ifglsused{#1}{%
    \glsentryshort{#1}%
  }{%
    \glsentrylong{#1}%
  }%
}

\newacronym{v2n}{V2N}{Vehicle To Network}
\newacronym{v2n2v}{V2N2V}{Vehicle to Network to Vehicle}
\newacronym{fdd}{FDD}{Frequency Division Duplexing}
\newacronym{cots}{COTS}{Commercial Off-The-Shelf}
\newacronym{dme}{DME}{Dedicated Measurement Equipment}
\newacronym{cgan}{CGAN}{Conditional Generative Adversarial Network}
\newacronym{dtpc}{DT-PC}{Digital Twin for Protocol and Connectivity}
\newacronym{jcas}{JCAS}{joint communication and sensing}
\newacronym{aql}{AQL}{Action-conditioned Q-Learning}
\newacronym{ibn}{IbN}{Intent-Based Networking}
\newacronym{imf}{IMF}{Intent Management Function}
\newacronym{rdf}{RDF}{Resource Description Format}
\newacronym{iri}{IRI}{Internationalized Resource Identifier}

\newacronym{rbac}{RBAC}{Role-Based Access Control}
\newacronym{pbac}{PBAC}{Policy-Based Access Control}
\newacronym{abac}{ABAC}{Attribute-based Access Control}
\newacronym{relbac}{RelBAC}{Relational/predicate-based Access Control}
\newacronym{zta}{ZTA}{Zero-trust architecture}
\newacronym{tls}{TLS}{Transport Layer Security}

\title{Authorization of Knowledge-base Agents in an Intent-based Management Function}

\author{\IEEEauthorblockN{Loay Abdelrazek}
\IEEEauthorblockA{\textit{Standards \& Technology} \\
\textit{Ericsson}\\
Sweden \\
first.last@ericsson.com}
\and
\IEEEauthorblockN{Leyli Karaçay}
\IEEEauthorblockA{\textit{Ericsson Research} \\
\textit{Ericsson}\\
Türkiye \\
first.last@ericsson.com}
\and
\IEEEauthorblockN{Marin Orli\'{c}}
\IEEEauthorblockA{\textit{Ericsson Research} \\
\textit{Ericsson}\\
Sweden \\
first.last@ericsson.com}
}

\maketitle

\begin{abstract}
As networks move toward the next-generation 6G, Intent-based Management (IbM) systems are increasingly adopted to simplify and automate network management by translating high-level intents into low-level configurations. Within these systems, agents play a critical role in monitoring current state of the network, gathering data, and enforcing actions across the network to fulfill the intent. However, ensuring secure and fine-grained authorization of agents remains a significant challenge, especially in dynamic and multi-tenant environments. Traditional models such as Role-Based Access Control (RBAC), Attribute-Based Access Control (ABAC) and Relational-Based Access Control (RelBAC) often lack the flexibility to accommodate the evolving context and granularity required by intent-based operations. In this paper, we propose an enhanced authorization framework that integrates contextual and functional attributes with agent roles to achieve dynamic, policy-driven access control. By analyzing agent functionalities, our approach ensures that agents are granted only the minimal necessary privileges towards knowledge graphs. 

\end{abstract}


\section{Introduction}
The evolution from 4G to 5G, and the upcoming transition to 6G, introduces increasingly complex network infrastructures that require new approaches to management and orchestration. In this context, \gls{ibn} has emerged as a transformative approach to network management, offering a more intuitive and automated way to configure, monitor, and optimize network resources. In a traditional network management system, administrators manually configure network elements using detailed, low-level commands, which can be cumbersome and prone to human error. In contrast, an \gls{ibn} allows operators to define high-level business or operational intents, such as "optimize network performance for high-priority traffic" or "ensure low latency for specific applications," without needing to specify the exact configuration steps. The system then translates these high-level intents into detailed network configurations and automatically enforces them across the network infrastructure.
By leveraging \gls{ai}, \gls{ml}, and advanced automation, intent-based management systems will be instrumental in simplifying the complexity of next-generation networks, improving operational efficiency, and ensuring optimal performance across increasingly diverse and dynamic services.

In \gls{imf} architecture we follow in this work, each IMF has a \gls{kb} that stores knowledge and facts for example the intent expectations and conditions, current state of intent, domain models, expert rules, etc. The knowledge graphs in the KB can be fully accessed by the agents inside the \gls{imf} whose tasks is to fulfill an intent~\cite{loops2022}. In the current architecture of the IMF, agents are assumed to belong to a single vendor and IMF is a black box function. However, future evolvement of IMF architecture lead to a multi-vendor architecture, where agents can be implemented by multiple vendors and separated logically in different parts of the network leveraging the opennes of the architecture for example the O-RAN architecture \cite{oranarch}.

In such architectural scenario, the method of accessing the \gls{kb} will need to be reconsidered. Rather than providing unrestricted access to the entire \gls{kb}, it will be necessary to implement a more granular access control model. This model would ensure that agents can only access the specific data and knowledge relevant to their role and functionality. In this paper, we propose an authorization approach that not only enhance agent's access management to the \gls{kb}, but also improve system efficiency by dynamically creating the authorization profiles based on agent's functionality and continously update the authorization profiles. Furthermore, restricting access to the full \gls{kb} could provide significant security benefits, especially in the event that an adversary manages to compromise one of the agents. If an agent is compromised, limiting its access to only certain graphs in the \gls{kb}, or specific facts that are relevant to its specific role, can minimize the damage and reduce the scope of information that the attacker can exploit.

Our contributions through this work, discussed in \ref{sec:authzapproach},  are as follows:
\begin{itemize}
	\item Hybrid authorization approach combining different concepts to ensure proper granularity of agent’s authorization to knowledge graphs.
	\item Dynamic authorization profiles for agents, created during runtime after the authentication process of an agent and during the bootstrapping and registration.
    \item Ensuring continuous authorization, enabling zero-trust principles in intent-based networks and granular authorization policies.
    \item Leveraging logical reasoning, allowing us to implement the authorization capability natively part of the logical reasoner to simplify integration and scalability.
\end{itemize}


\begin{figure}
	\centering
	\includegraphics[width=0.8\linewidth]{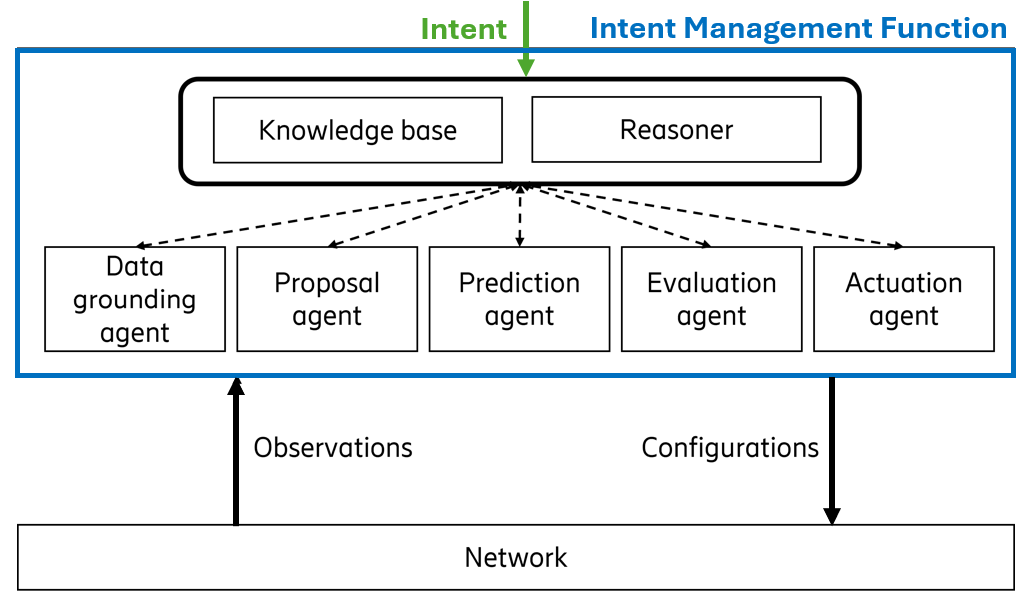}
	\caption{Intent-based Management System Framework}
	\label{fig:imf}
\end{figure}

\section {Background}

\subsection {Knowledge-based Agent}
A standardized view or standardized definition of what constitutes an agent is not widely agreed upon. This paper relies on the scientific definition of an agent found in~\cite{russell2016artificial}, where it states that an agent is anything that can be viewed as perceiving its environment through sensors and acting upon that environment through actuators.

There are multiple types of agents, for example learning agents, simple reflex agent, knowledge-based agent and others. The nearest type of agent in terms of functionality to the intent-based management function to the best of our knowledge are the knowledge-based agents (i.e reasoning agents). The main components to support reasoning agents are the \gls{kb} where facts and knowledge graphs are represented using a knowledge representation language (e.g, \gls{rdf}\cite{rdf}) and a reasoner that implements logical rules used for inference to proof goals, derive facts and conclusions. The nature of this types of agents are dynamic, meaning that they follow an autonomous loop where they continuously observe and update the \gls{kb} or query the \gls{kb} for new facts where they can act upon. The assumption is that the agent is deployed with an understanding of its functionality that is submitted upon agent registration to the \gls{imf}.

The agents can have different possible roles as illustrated in Figure \ref{fig:imf}, for example data grounding agents consistently monitor the underlying system and refresh the knowledge in the IMF. If any of the system's intents are no longer met, the proposal agents are activated. They suggest a range of actions that could potentially resolve the issue and achieve the desired state. Prediction agents estimate the impact of each proposed action on the system. Then, the evaluation agent evaluates the suggested actions based on these predictions and identifies the most viable option. Finally, the actuation agent implements the chosen action within the network.

\subsection {Intent Management Function}
As defined by the TM Forum~\cite{tmf1253}, intent refers to the \emph{formal specification of all expectations, including requirements, goals, and constraints provided to a technical system}. Natural language and other domain-specific languages can be utilized to define an intent, but they need to be interpreted and translated into a unified intent model.

The \gls{imf}, as outlined by the TM Forum, is responsible for managing these intents and executing intent-driven operations. The structure of the \gls{imf} is illustrated in Figure~\ref{fig:imf}. The core framework of the \gls{imf} consists of a \gls{kb} and a reasoner. The \gls{kb} stores information including the network state, intent life cycle, expectations, and intent ontology. Reasoner coordinates the closed loop operations to handle the intents and executes the corresponding set of agents according to their roles.

While our approach benefits from using the reasoner within the \gls{imf} implementation, our method may be applied to other implementation architectures utilizing agents to abstract and implement parts of the intent-handling loop.

\subsection {Authorization Principles}
The authorization method adopted in this work to authorize the agents towards the \gls{imf} was defined by following widely known and good practice authorization principles. OWASP~\cite{owasp} defines several principles for implementing authorization logic that is robust, context-aware, and scalable. Moreover, the principles of Zero Trust Architecture (ZTA) are being intensively studied within the mobile network ecosystem across various standards development organizations, for example 3GPP \cite{3GPP-33894}, as well across mobile network technology providers~\cite{olsson20215g}. Therefore, adopting ZTA principles in emerging paradigms and architectures within the mobile network domain becomes essential.

ZTA principles has been defined by NIST~\cite{stafford2020zero}, primarily tenets~1, 3 and 6 are those that this work aligns with. \textbf{Tenet 1}: presents that all data sources and computing services are considered resources -- in this work agents as well as the \gls{kb} that holds the graphs are considered resources. \textbf{Tenet 3}: states that access authorization should be per session and follow a least priviledge approach, meaning that authorization to one resource does not automatically grant access to another resource. In this work, different levels of granularity for authorization is applied on the \gls{kb} resources, in addition inherited authorization of agent is prohibited. Finally \textbf{Tenet 6}: states that all resource authentication and authorization is dynamic and enforced before access is allowed. In this work, authorization is applied upon each query provided by the agent.

The authorization principles followed in this paper can be summarized as follows:

\begin{itemize}
  \item Agents and knowledge base are resources.
  \item Least privilege authorization of agents.
  \item Ensure permissions with every request (i.e query or assert).
  \item Access is denied by default.
  \item Permissions shall not be inherited.
  \item Authorization policy is dynamically updated.
\end{itemize}

In this work, we assume that an authentication method is available to authenticate the agents with the \gls{imf}. The Agent authentication can possibly adopt existing mechanisms made available by the \gls{tls} protocol. The \gls{imf} and the agent perform mutual authentication when connection is established. For that purpose, both the \gls{imf} and the agent are required to have valid certificates, and the verification follows the standard \gls{tls} procedure: after the agent connects and verifies the \gls{imf}'s certificate, the agent is required to present its own certificate before the \gls{imf} grants access. In addition to requiring the agent to present a valid certificate, we also require that the subject field in the certificate contains a \emph{role} annotation, for example \texttt{Subject: CN=agent, role=grounder}. Upon successful connection, the agent's role is stored in the authorization graph and used for authorization as given in Figure~\ref{fig:authz_graph}.

\section{Related Work}
While there has been extensive research on storing, representing, and reasoning with RDF knowledge, studies focused on security and access control for RDF stores are relatively limited~\cite{reddivari2007policy}. The most widely used method is \gls{rbac}~\cite{guiri1995new}, where agents are assigned specific roles, and permissions are granted based on those roles to determine what actions they can perform.
However, one limitation of \gls{rbac} is that the roles are predefined, and any change in an entity’s access rights requires modifying its role, making the system rigid. Another limitation in \gls{rbac} is inheritance of access rights, as the current application of authorization is used in the application of social network graphs, where inheritance of access to resources in the graph is reasonated. However, in \gls{ibn} the agents should have a limited view on what shall be accessed, additionally one role could be divided into sub-roles depending on the funcitonality of the agent, where each will have its own part of the graph that requires to access.

In~\cite{cirio2007role} \gls{abac} has been explored where contextual attributes are leveraged, allowing for dynamic role assignments based on user credentials and environmental factors. This approach offers greater flexibility compared to traditional \gls{rbac}, where roles are typically assigned statically by security administrators, and \gls{rdf} is used to model and represent data about users, roles, attributes, and access permissions in a machine-readable format, yet again the limiation of access rights inheritance.

In another method that can be used is \gls{pbac}, which is proposed by~\cite{reddivari2007policy}. It addresses the need for robust access control mechanisms in \gls{rdf} stores, which are integral to many semantic web applications. In this model, agents submit requests to perform actions (e.g. deleting a triple) on the \gls{rdf} store, and the system determines whether to permit or deny these actions based on predefined policies. These policies are defined by a collection of rules that consider factors such as the requesting agent's attributes, the type of action, the history of previous actions, and the contents of the store. An example of a policy rule can be: \textit{Only agents assigned to an editor role are allowed to
insert or delete triples}.


The work done in \cite{giunchiglia2008relbac} and \cite{clark2022relog} presents permissions as relations between users and objects in the knowledge-graph. The core essence of the \gls{relbac} approach is to simplify the management of an expanding number of users and memberships allowing for propagation of permissions, thus allowing the same permissions for all users that are members of a group, which is yet another can be foreseen as a limitation as it allows inheritance of permissions.

Social networks has been the main use case study in different work, where access controls are applied on the graphs constituting the network. The main objective besides offering authorization is to manage the expanding number of users, thus the common limitation in most of the work on knowledge-graph access control is the inheritance and propagation of permissions among different users that may share the same role or members of the same group. The nature and types of knowledge-graphs in \gls{ibn} is slightly different and requires applying stricter and more granular access control on the knowledge graphs.

Moreover, the proposed control mechanisms in \cite{rebecchi2024revealing} is to protect the \gls{imf} by ensuring authentication and authorization mechanisms. However the proposed control considers authentication and authorization between the different \gls{imf} in the network that are communicating with each other over the intent APIs and not considering the agents that are part of the cognitive loop.
\section{Threat model in Intent-based Management} \label{sec:threats}

The \gls{ibn} threat model has been explored in different work \cite{rebecchi2024revealing} and \cite{benzaid2020zsm}, where both works discussed the possible threats to intent-based management but from an architectural point of view, whether as threats on ETSI Zero-touch and Service Management (ZSM) architecture or the O-RAN architecture. Both works considered attack surfaces that are related to the architecture and different interfaces, for example threats that are related to improper validation of intents, rogue \gls{imf} in the network, and AI-related threats that may impact the closed loop.

On the other hand, the agents that are responsbile to handle the autonomous loop play a significant role in fulfilling the intent's goals, where currently they have a wide access over all of the different objects in the \gls{kb}. In the aforementioned previous work, the assumed trust boundary was the IMF itself, meaning that anything external to the \gls{imf} is not trusted and assumed to be a threat surface. This approach is essential but it does not take into consideration the possible different types of deployments of \gls{imf}.

In this work, we consider \gls{imf} as a logical functional component composed of several subcomponents, including the agents, the reasoner, and the knowledge base where the \gls{kb} is the asset to be protected. Treating the \gls{imf} as a logical entity, results in the possibility to redefine the trust boundary in \gls{ibn} to be the subcomponents of an \gls{imf} which implies that each agent is its own trust boundary, the \gls{kb} and reasoner forms another trust boundary.




\section {Agents Authorization Mechanism} \label{sec:authorization}

Authorization is one form of access control that is defined according to~\cite{nistauthz} as \emph{The access privileges or a set of permissions granted to a user or a system entity to access a system process and act according to the granted privileges}. Agents in a cognitive loop should be authorized to only be able to view (i.e query) or update (i.e assert) facts based on their functionality and affiliation to the problem to be solved. In other words, any agent in an automation loop should not be able to act on goals or facts that should be under the responsibility of another agent that is part of the loop.  Additionally, agent's authorization becomes crucial to aid on preventing the aforementioned threats in \ref{sec:threats} as well to accommodate to any type of future deployments.

The knowledge facts the agents are responsible to view or update are tightly coupled to the knowledge graphs present in the \gls{kb}. Depending on the use case, the agents may require to have a view of an entire graph, just a part of the graph, or even view specific triples in the knowledge graph. Hence, authorization methods require flexibility and different granularity levels to accomodate the dynamic nature of agents. This will require that authorization can be done on graph level, sub-graph level, and relationship (i.e predicate) level.

In the following sections we present the proposed authorization approach, the ontology, and examples of authorization inference rules that are used to create the dynamic authorization policies, assign permissions and act according to the resulted authorization state.

\subsection{Hybrid Authorization Approach} \label{sec:authzapproach}

\begin{figure}[!ht]
	\centering
	\includegraphics[width=0.95\linewidth]{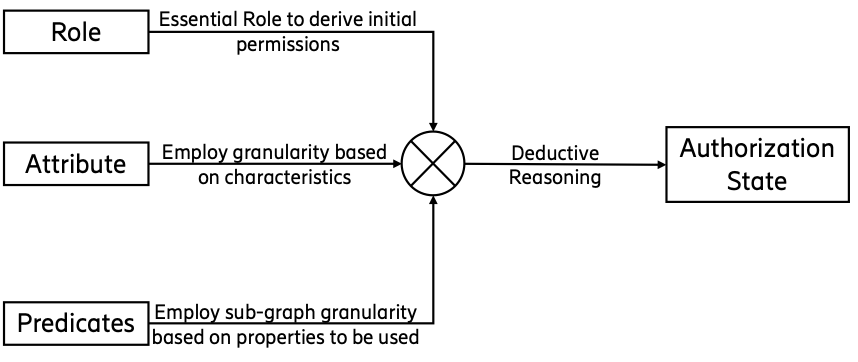}
	\caption{A Hybrid Authorization Approach.}
	\label{fig:authzApproach}
\end{figure}

In this work, we adopted a hybrid authorization approach, where we combined role-, attribute-, and predicate-based methods. The inference result of each method are conjuncted to derive a conclusion around the authorization state as it is illustrated in Figure~\ref{fig:authzApproach}. Each method is used to derive a set of facts that either used as an input to our hybrid authorization approach.

In our approach, the facts and knowledge obtained from three aforementioned methods are intended to dynamically build authorization profile for the agent. Part of the authorization is to prepare an authorization graph for different roles. Then, authorization profile is used during the operations of the agent to match each incoming request against the facts present in the agent authorization graph. Following information is part of the authorization profile:

\begin{itemize}
    \item Identity of the Agent (i.e its registered \gls{iri}).
    \item The resources an agent is responsible to handle (i.e resources to be accessed in the knowledge graphs).
    \item The authorization role.
    \item Expected permission.
    \item Authorization and authentication state.
    \item Authorization attributes (e.g, the intent object).

\end{itemize}

\gls{rbac} is a widely known access control method, but it does not provide a granular method to ensure that an agent can only access a limited view, since the functionality of the agent can be divided into several sub-functionalities. For example, a grounding agent can have different sub-functionalities like \gls{ue} grounding, network topology grounding, power consumption grounding and others depending on the data to be collected.
Thus, aggregating all these sub-roles into one Grounding role, provides an excessive access. Nevertheless, since we can have limitless number of sub-roles as it depends on the data being collected, there can not be a one-to-one mapping as we cannot foresee in advance what would be the different sub-role of an agent or the use case across different intents.
Therefore, using \gls{rbac} as an initial step defines the main role of the agent, for example Grounding, where this fact is used to limit the initial set of permission scope of this agent type. 

In the scope of agents fulfilling an intent, contextual attributes are leveraged, for example the agent attributes could be the intent object that they will handle, the type of the intent they are responsible for (e.g, service intent, business intent, or resource intent), the IMF domain, the metrics they are collecting and updating, issues that they are responsbile to evaluate, or values that are in their responsbility to predict among other attributes that can provide more context of the agent's capability.

Relationships have been proposed in multiple works as discussed earlier. This work uses the concept of relationships but with the focus on the predicates that connect two nodes in a graph itself. The predicate used to define the relationship between the agent as a subject and another resource in the \gls{kb} as an object can be used to allow even more granular access in the knowledge graph based only on the defined relationship (Agent, Predicate, Object). Even if the predicate is being used to define a relationship between another agent and another object instance, both agents will not be able to access the same object.

\subsection{Authorization Flow} \label{sec:authzflow}
\begin{figure}[!ht]
	\centering
	\includegraphics[width=1\linewidth]{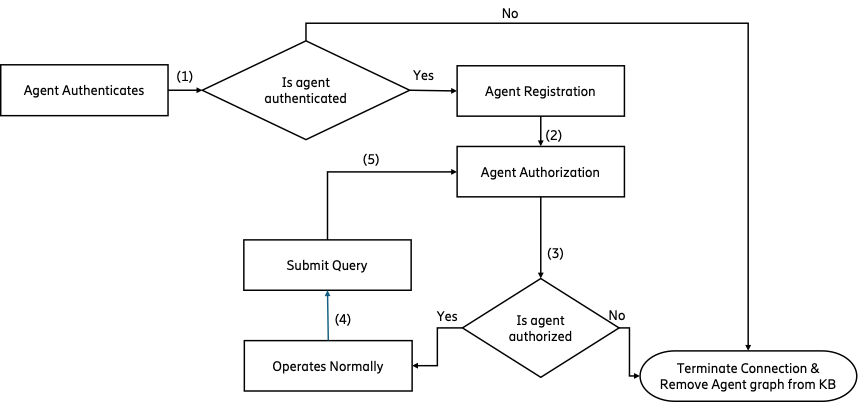}
	\caption{Authorization Flow}
	\label{fig:authzflow}
\end{figure}

The authorization of an agent occurs at different stages during the lifecycle of the agent, as illustrated in Figure \ref{fig:authzflow}. The authorization occurs after the agent is authenticated to the reasoner and each time the agent is submitting a query to the \gls{kb}.

If the agent is authenticated, then it registers itself and provides the reasoner with different information, for example where to find its callback handler, its identity, the type of predicates and resources it requires access to, etc. Otherwise, if the agent is not authenticated, the session gets terminated and any information related to the agent is retracted from the \gls{kb}.

The authorization inference rules check the submitted facts and match them against the agent's authorization profile to allow or block the operation. This happens when the agent is invoked by the reasoner to run a function and returning the results, or if the agent is submitting a query to retrieve or update facts. If agent is authorized, it operates normally and the query is allowed. Otherwise, the session is terminated and the agent's graph and related objects are retracted from the \gls{kb}.

\subsection{Authorization Ontology} \label{sec:authzontology}

The intent-based management paradigm follows a model-driven architecture where the world is viewed as an abstraction model that is loaded into a \gls{kb}. The description of the concepts, facts and relationships that can exist for a domain or a problem to be solved is defined in an ontology. In this paper we used the description of authorization concepts and controls that was modeled as part of an overall security ontology in TM Forum~\cite{TMFSecurity}. The authorization ontology is composed of nodes that represent the authorization concepts defined as classes and edges are the predicates or the relationships between the different nodes. The ontology currently defines 26 concept nodes that have 30 edges defining the relationships between them.


The authorization ontology describes the concepts of authorization policy, roles, attributes and other knowledge needed to build the agents authorization profile graphs. Examples of knowledge that can be derived leveraging the authorization ontology are as follows:
\begin{itemize}
	\item (Agent, hasRole, Role),
	\item (Agent, hasAuthorizationProfile, AuthorizationProfile),
	\item (AuthorizationProfile, authorizedPredicates, Relations),
	\item (AuthorizationProfile, accessTo, TargetResource),
	\item (AuthorizationProfile, allowedPermission, Permission),
	\item (AuthorizationProfile, allowedPermissionValues, Actions),
	\item (AuthorizationProfile, accessGranted, Decision).
\end{itemize}

Authorization knowledge can be divided into two types or categories. The first category relates to \textbf{static knowledge}, which is assumed to be known beforehand, possibly during the development and deployment phases of an agent. This type of knowledge constitutes of facts such as the functionality of an agent, which intent type or problem type it shall be associated with, etc. These facts can then be used to derive further knowledge and produce new facts as a result.

The second category of knowledge is \textbf{dynamic knowledge}, which is the type of knowledge that is acquired during agent operation. The dynamic knowledge constitutes of facts that create the actual authorization policy of an agent. Moreover, dynamic knowledge will also be responsible on updating the authorization policy if needed. For example, a Grounding agent submits its capabilities that indicate which data it is responsible for collecting and monitoring, implying in turn which triples should be matched during authorization inference.

\subsection{Authorization Inference} \label{sec:authzinference}

The authorization rules are based on a declarative language. In this paper, we used \gls{rdf} serialized in Turtle~\cite{turtle} to build the inference rules. There are different rules that derive several information and alter the knowledge graphs for the agents accordingly.

Authorization of an agent goes through different phases. Firstly, after the agent is authenticated using its certificate, the agent's role is defined as part of the attributes in the certificate. The authorization rule assigns an initial profile related to the functionality that is derived from the agent's role. The initial authorization profile includes basic information, for example the assigned default permissions for this type of agent, or any other information that is assigned to all the agents with same role. Then the rule is responsible to assert all of this information and create a graph in the \gls{kb} that holds the authorization policy of the agent. Each agent has its own profile dynamically created and stored as a graph in the \gls{kb}. This authorization graph acts as a policy that is matched and checked with each new operational request (i.e. update of facts).

In the authorization process, the reasoner part acts as the Policy Decision Point (PDP) and the Policy Enforcement Point (PEP) for the \gls{imf}. Within the authorization process itself there are different steps taken as follow:
\begin{enumerate}
	\item During registration, the agent submits the name of the resources that it will handle, in the form of graph triples, where the agent is the subject, the relationship is the predicate, and the resource that it requires access to is the object. From the triple, the authorization rules can derive which graphs in the \gls{kb} the agent should be confined to and only allowed access to.
	\item The reasoner equipped with the authorization inference rules, derives the initial facts about the agent and create an authorization profile graph for agent.
	\item The authorization profile graph is then used during the operations of the agent to match each incoming query against the facts present in the agent authorization graph.

\end{enumerate}


\begin{figure}[!ht]
	\centering
\begin{lstlisting}[language=Turtle]
@prefix  ex: <https://www.example.com/> .
@prefix imf: <https://www.example.com/imf/> .
@prefix sec: <https://www.example.com/security/> .


ex:UpGroundingAgent
  a imf:Agent ;
  sec:hasAuthorizationProfile sec:GroundingAgentProfile;
  sec:consumerIdentity <http://grounding-agent.example.com>;
  sec:hasAuthorizationRole sec:Grounder ;
  sec:accessTo [ sec:object ex:Latency ;
                 sec:target ex:gNB01 ;
               ];
  sec:allowedPermission [ ex:Read ex:Write ];
  sec:allowedPermissionValues [ sec:readPermission ex:Match;
                                sec:writePermission ex:Assert ];
  sec:hasAuthorizationAttribute ex:LatencyIntent .
\end{lstlisting}
	\caption{Creating an authorization profile graph for the agent serialized in Turtle}
	\label{fig:authz_graph}
\end{figure}

\begin{figure}[!ht]
	\centering
	\includegraphics[width=1\linewidth]{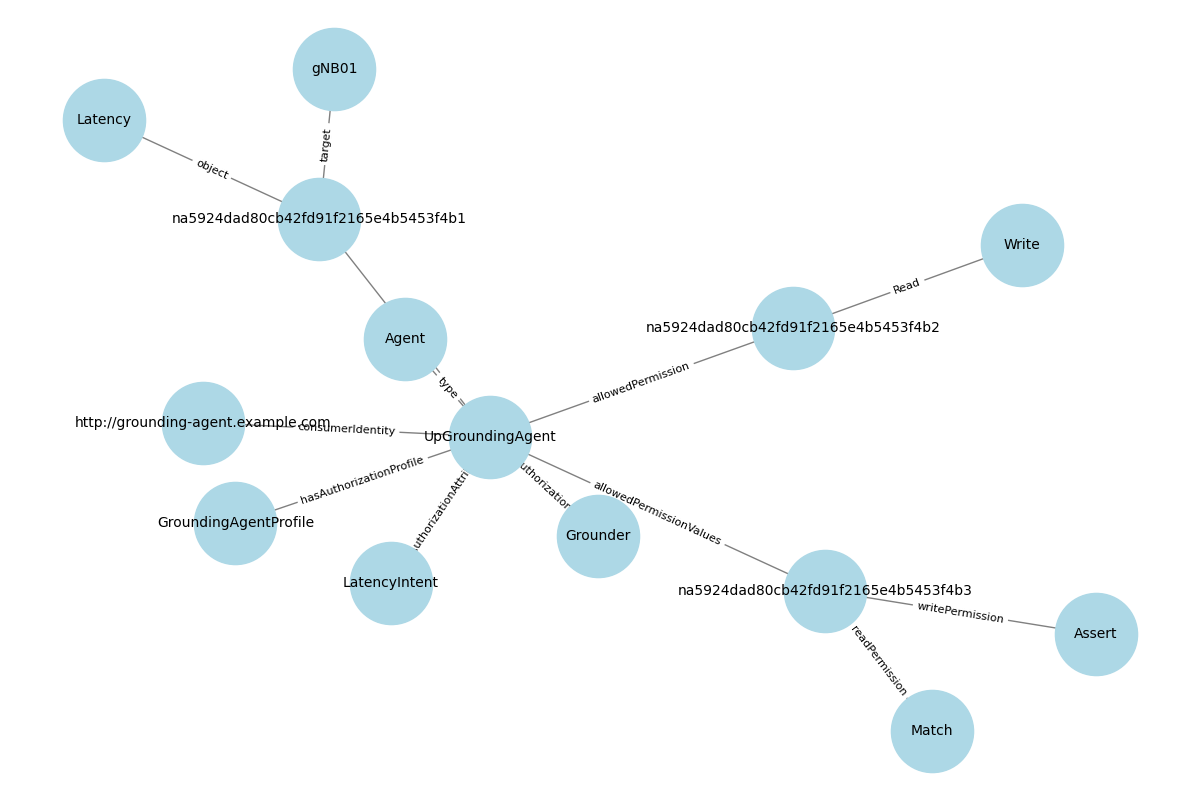}
	\caption{Graph representation of a grounding agent authorization profile }
	\label{fig:authz_profile_graph}
\end{figure}

Figure~\ref{fig:authz_graph} illustrates an example of the authorization graph of an agent serialized in Turtle. It typically includes a parent authorization profile it that is a member of, the agent's identity (i.e can be its handler IRI, FQDN, etc..), the role it has, the resources in the \gls{kb} (i.e nodes in a graph) that it is expected to have access to, the allowed permission actions and if there are specific values or commands that are used part of the permission, and finally any other attributes that can be used to add more context to the agent, for example which intent that is in its scope. The graph representation of the authorization profile is represented in Figure~\ref{fig:authz_profile_graph}.

As mentioned in the authorization flow, the first step after the agent's authentication and registration is to derive its role from the available facts, and check if the role is a known one, what is the role's identity, associated functionality, etc. The second step is to create the agent's own authorization profile graph that will be used to hold its security context. The logical formulation below is one example to represent how to derive the role from knowing the functionality of the agent. The statement in \eqref{equ:1} can be read as that for all authenticated and registered agents in the \gls{kb} there exists a capability and an authorization role, such that the capability associated with the agent implies that the role of an agent (A) is (R).
\begin{equation}
	\forall A \exists F \exists R \ {AgentFunction(A,F)} \Rightarrow {Role(A,R)}
	\label{equ:1}
\end{equation}

Sequentially, from knowing the role, the next step is to derive what are the possible permissions that can be initially associated with the role, if there are no other rules that state otherwise. A possible formulation could be as follows:
\begin{multline}
	\forall A \exists R \exists P \ {AgentRole(A,R)} \land {InitialPermission(R,P)} \\ \land \neg {Exceptions(A,R,P)}  \Rightarrow {Permission(A,P)}
	\label{equ:2}
\end{multline}

The statement in~\eqref{equ:2} states that for all the agents having their roles identified, these roles have some initial permissions (P) and there is no constraints or exceptions found for the agent with that specific role. An exception could be additional inference rules that are targeting the sub-functionality of the role. For example, a \gls{ue} grounding agent could have a different set of permissions than the topology grounding or the rest of all other agents with role grounding. An authorization profile graph is finally created with the facts submitted by the agent and the derived facts indicating its scope, as illustrated in Figure~\ref{fig:authz_graph}.
A graph is composed of triples $(Subject,Predicate,Object)$, where the agent is the subject, predicate is related to a property or a relationship to an Object, and the object is another node in the graph. For example, \textit{(ex:UpGroundingAgent,ex:accessTo,ex:ListOfResources)}, means that the agent which is of type user plane grounding agent that is responsbile to ground data, for example throughput ond latency,is an instance of a subject will have access to the specified elements part of the resource lists. The resources list is an object that defines the target or the network functions it is responsible to collect data for and the type of metric it will observe and update which is latency in this example. The object represented in the \textit{ex:ListOfResources} could be one of the facts that are submitted by the agent during a registration flow. An object with no assigned value is equivalent to a wildcard, which implies that an agent will monitor any type of data or KPI on all possible nodes and update the knowledge graphs accordingly. To ensure a level of least privilege, an initial check is done that the agent priviledges refer to actual values of object and wildcard instances which will lead to an excessive access. An example of the logical statement is provided in equation~\eqref{equ:3}:

\begin{multline}
	{LeastPrivilege(Agent,(Subject,Predicate,Object))} \Leftarrow \\
	\neg{WildCardValue(Object)}
	\label{equ:3}
\end{multline}

The above statement~\eqref{equ:3} just checks that the object part of all the triples provided by the agent as part of their registration or facts part of their knowledge graph in the \gls{kb}, referring to the resource to be accessed, has a value and not a wildcard variable. The authorization profile graph of an agent will be checked on every request (i.e query).

\begin{multline}
	{AuthorizationState(Agent,Request)} \Leftarrow \\
	{ExtractRequest(Request,PatternR)} \land \\
	{PatternLookup(Agent,PatternK, AgentGraph)} \land \\
	{Match(PatternR,PatternK)}
	\label{equ:4}
\end{multline}

Finally, the statement in~\eqref{equ:4} at each request initiated by the agent the resources the agent would update or request to read, then it searches for the known patterns in the agent graph that the agent is allowed to access and if there are no new facts. If both patterns match, then agent is authorized to perform operations on the resources submitted in the request.

\section{Conclusion} \label{sec:conclusion}
In this work, we proposed an authorization method that is capable of authorizing knowledge-base agents that are part of an \gls{imf}. The authorization method leveraged logical inference rules that follow a hybrid approach that can provide a proper level of granularity and ensure agents have the least privilege access on the resources within the knowledge base and avoid the common limitation of authorization inheritance in knowledge graphs. This approach not only addresses the limitations of traditional models like \gls{rbac}, \gls{abac} and \gls{relbac} by avoiding access inheritance among agents having the same role, but also simplifies the management of authorization profiles by creating dynamic authorization profiles during runtime of authenticated agents. As a result, the proposed framework offers a practical and scalable solution for securing agent's behavior in modern, automated network infrastructures.

Additionally, we foresee that standardization is an enabler to ensure the security of \gls{imf} architecture, by requiring that \gls{imf} should be equipped with the authorization capabilities proposed in this work to enforce access control on agents, an authorization ontology that is unified across the different intents Standard Developemnt Ogranization (e.g, 3GPP, O-RAN, TM Forum) is needed to ensure interoperability between multi-vendor components, and finally the supported security capabilities of the \gls{imf}, for example agent's authorization, to be part of the IMF capability profile used for registration and discovery of an \gls{imf}.

\section*{Acknowledgement}
This work was supported by the Scientific and Technological Research Council of Turkey (TUBITAK) through the 1515 Frontier Research and Development Laboratories Support Program under Project 5169902.

\bibliographystyle{IEEEtran}
\bibliography{ref}

\begin{thebibliography}{10}
\providecommand{\url}[1]{#1}
\csname url@samestyle\endcsname
\providecommand{\newblock}{\relax}
\providecommand{\bibinfo}[2]{#2}
\providecommand{\BIBentrySTDinterwordspacing}{\spaceskip=0pt\relax}
\providecommand{\BIBentryALTinterwordstretchfactor}{4}
\providecommand{\BIBentryALTinterwordspacing}{\spaceskip=\fontdimen2\font plus
\BIBentryALTinterwordstretchfactor\fontdimen3\font minus \fontdimen4\font\relax}
\providecommand{\BIBforeignlanguage}[2]{{%
\expandafter\ifx\csname l@#1\endcsname\relax
\typeout{** WARNING: IEEEtran.bst: No hyphenation pattern has been}%
\typeout{** loaded for the language `#1'. Using the pattern for}%
\typeout{** the default language instead.}%
\else
\language=\csname l@#1\endcsname
\fi
#2}}
\providecommand{\BIBdecl}{\relax}
\BIBdecl

\bibitem{loops2022}
A.~C. Baktir, A.~D.~N. Junior, A.~Zahemszky, A.~Likhyani, D.~A. Temesgene, D.~Roeland, E.~D. Biyar, R.~F. Ustok, M.~Orlić, and M.~D’Angelo, ``Intent-based cognitive closed-loop management with built-in conflict handling,'' in \emph{2022 IEEE 8th International Conference on Network Softwarization (NetSoft)}, 2022, pp. 73--78.

\bibitem{oranarch}
{O-RAN}, ``O-ran architecture description 13.0,'' 2025.

\bibitem{russell2016artificial}
S.~J. Russell and P.~Norvig, \emph{Artificial intelligence: a modern approach}.\hskip 1em plus 0.5em minus 0.4em\relax pearson, 2016.

\bibitem{rdf}
\BIBentryALTinterwordspacing
W3C, ``Rdf 1.2 concepts and abstract syntax,'' 2025. [Online]. Available: \url{https://www.w3.org/TR/rdf12-concepts/}
\BIBentrySTDinterwordspacing

\bibitem{tmf1253}
{TM Forum}, ``{IG1253 Intent in Autonomous Networks},'' 2022.

\bibitem{owasp}
{OWASP}, ``{OWASP Authorization Principles}.''

\bibitem{3GPP-33894}
{3GPP}, ``{3GPP 33.894} study on applicability of the zero trust security principles in mobile networks, release 18.''

\bibitem{olsson20215g}
J.~Olsson, A.~Shorov, L.~Abdelrazek, and J.~Whitefield, ``5g zero trust--a zero-trust architecture for telecom,'' \emph{Ericsson Technology Review}, vol. 2021, no.~5, pp. 2--11, 2021.

\bibitem{stafford2020zero}
V.~Stafford, ``Zero trust architecture,'' \emph{NIST special publication}, vol. 800, no. 207, pp. 800--207, 2020.

\bibitem{reddivari2007policy}
P.~Reddivari, T.~Finin, A.~Joshi \emph{et~al.}, ``Policy-based access control for an rdf store,'' in \emph{Proceedings of the IJCAI-07 workshop on semantic web for collaborative knowledge acquisition}, 2007.

\bibitem{guiri1995new}
L.~Guiri, ``A new model for role-based access control,'' in \emph{Proceedings of 11th Annual Computer Security Application Conference}, 1995, pp. 249--255.

\bibitem{cirio2007role}
L.~Cirio, I.~F. Cruz, and R.~Tamassia, ``A role and attribute based access control system using semantic web technologies,'' in \emph{OTM Confederated International Conferences" On the Move to Meaningful Internet Systems"}.\hskip 1em plus 0.5em minus 0.4em\relax Springer, 2007, pp. 1256--1266.

\bibitem{giunchiglia2008relbac}
F.~Giunchiglia, R.~Zhang, and B.~Crispo, ``Relbac: Relation based access control,'' in \emph{2008 Fourth International Conference on Semantics, Knowledge and Grid}.\hskip 1em plus 0.5em minus 0.4em\relax IEEE, 2008, pp. 3--11.

\bibitem{clark2022relog}
S.~Clark, N.~Yakovets, G.~Fletcher, and N.~Zannone, ``Relog: a unified framework for relationship-based access control over graph databases,'' in \emph{IFIP Annual Conference on Data and Applications Security and Privacy}.\hskip 1em plus 0.5em minus 0.4em\relax Springer, 2022, pp. 303--315.

\bibitem{rebecchi2024revealing}
F.~Rebecchi, D.~Cho, L.~Abdelrazek, H.~Forssell, and J.~Olsson, ``Revealing the threat landscape of intent-based management in o-ran,'' in \emph{2024 27th Conference on Innovation in Clouds, Internet and Networks (ICIN)}.\hskip 1em plus 0.5em minus 0.4em\relax IEEE, 2024, pp. 106--113.

\bibitem{benzaid2020zsm}
C.~Benzaid and T.~Taleb, ``Zsm security: Threat surface and best practices,'' \emph{IEEE Network}, vol.~34, no.~3, pp. 124--133, 2020.

\bibitem{nistauthz}
{NIST}, ``{NIST Special Publication 800-53A Revision 5},'' 2022.

\bibitem{TMFSecurity}
{TMforum}, ``Tr292i security ontology v.3.7.0,'' 2025.

\bibitem{turtle}
\BIBentryALTinterwordspacing
W3C, ``Terse rdf triple language,'' 2025. [Online]. Available: \url{https://www.w3.org/TR/rdf12-turtle/}
\BIBentrySTDinterwordspacing

\end{thebibliography}

\end{document}